\documentclass[10pt,final]{acmconf}
\usepackage {FIE}
\usepackage {epsfig}
\usepackage {amsmath}
\usepackage[dvips]{color}

\begin{document}
\title{Mapping DEVS Models onto UML Models}
\author{Dmitry~Zinoviev\\
  Department of Mathematics and Computer Science, Suffolk University\\ 
  32 Derne St., Boston, MA, 02114 USA\\
  Dmitry@MCS.Suffolk.EDU}
\maketitle

\noindent%
\vskip\baselineskip

\abstract{
Discrete event simulation specification (DEVS) is a formalism
designed to describe both discrete state and continuous state
systems. It is a powerful abstract mathematical notation. 
However,
until recently it lacked proper graphical representation,
which made computer simulation of DEVS models a challenging
issue. Unified modeling language (UML) is a multipurpose graphical
modeling language, a de-facto industrial modeling standard. There exist
several commercial and open-source UML editors and code
generators. Most of them can save UML models in XML-based XMI files
ready for further automated processing. In this paper, we propose a
mapping of DEVS models onto UML state and component diagrams. This
mapping may lead to an eventual unification of the two modeling
formalisms, combining the abstractness of DEVS and expressive power
and ``computer friendliness'' of the UML.
}

\section{INTRODUCTION}

Discrete event simulation specification
(DEVS~\cite{zeigler84,zeigler90,zeigler00}) is a powerful formalism
for describing discrete event state systems. Various flavors of this
formalism have been developed. In this paper, we are using classic
DEVS with ports, which we call, simply, DEVS.

Being hierarchical and encapsulated, the DEVS formalism can be
naturally implemented in an object-oriented language, such as
Java~\cite{sarjoughian04}. However, Java-based simulation environments
have never been standardized (unlike the DEVS formalism).

An important drawback of the DEVS formalism is the lack of a
standardized graphics representation. In~\cite{feng04}, an attempt has
been undertaken to develop DCharts, a graphics language for DEVS
models. DCharts is a UML-based language; however, it does not strictly
follow any UML standard. Moreover, the DEVS-to-DCharts transformation
proposed in~\cite{feng04} collapses all DEVS states into one DCharts
state, essentially eliminating the discrete state nature of DEVS
models.

In~\cite{hong04}, it is proposed that atomic DEVS models be
represented with the help of UML sequence diagrams. However, the
sequence diagrams show actual, rather than potential, flow of
events. Because of this, a sequence diagram is limited to a particular
scenario and cannot unambiguously describe the behavior of a system in
its entirety.

An excellent mapping between DEVS models and UML state charts has been
introduced in~\cite{schulz00}. However, the paper does not suggest a
formal mathematical way of constructing the state charts, and thus
avoids the issue of the structural clash between DEVS continuous
states and UML finite states. It also relies on the older versions of
the UML ($<2.0$), which did not have an explicit notion of time.

In this paper, we propose a consistent DEVS-to-UML mapping that takes
care of all issues mentioned above. Further restricting the mapping to
the executable UML (which is a proper subset of UML~\cite{mellor2002})
would make a seamless connection between a DEVS model and an UML
simulation process, but this topic is beyond the scope of this paper. 

\section{DEVS FORMALISM}

DEVS supports two complementary models of describing discrete systems:
an atomic model that specifies the behavior of an elementary system,
and a coupled model that allows us to form more complex models by 
structurally interconnecting atomic and other coupled models.

A DEVS atomic model is a state machine with input ports
$\mathrm{IP}=\{\pi^{in}_i\}$ and output ports
$\mathrm{OP}=\{\pi^{out}_i\}$. Events are associated with input and
output ports (they happen at input ports and are generated at output
ports). In general, states, unlike events, are not discrete.

\subsection{Atomic Models}
Atomic DEVS model $M_a$ is a tuple of nine values $M_a=\{\mathrm{IP},
\mathrm{OP}, X, \Sigma, Y, \delta_\mathrm{in}, \delta_\mathrm{ext},
\lambda, t_a \}$.  Here, $\mathrm{IP}$ and $\mathrm{OP}$ are sets of
input and output ports.

$\Sigma$ is a set of states $\{\sigma_i\}$. A DEVS state $\sigma_i$ is
uniquely identified with a set of state variables
$\gamma_i=\{\gamma_{ij}|\sigma_i\ne\sigma_k \Leftrightarrow \exists j:
\gamma_{ij} \ne \gamma_{kj}\}$.

$X$ is a set of input events $\{x_i=\left(\pi^{in}_i,v_i\right)\}$,
where $\pi^{in}_i\in \mathrm{IP}$ is the port name, and $v_i$ is the
event value.  Every event has an associated timestamp, or scheduled
time---the time when the event is triggered.

$Y$ is a set of output events \{$y_i=\left(\pi^{out}_i,v_i\right)\}$,
where $\pi^{out}_i\in \mathrm{OP}$ is the port name, or the event
type, and $v_i$ is the event value.
$\delta_\mathrm{int}\left(\sigma_i\right): \Sigma\rightarrow \Sigma$
is the internal transition function.
$\delta_\mathrm{ext}\left(\sigma_i,e_i,x_j\right):\left(\Sigma\times
R\times \mathrm{IP}\right)\rightarrow\Sigma$ is the external
transition function, where $e_i$ is the elapsed time in state
$\sigma_i$.  $\lambda\left(\sigma_i\right):\Sigma\rightarrow Y$ is the
output function.  $t_a\left(\sigma_i\right):\Sigma\rightarrow R$ is
the time advance function.

The semantics of the model are as follows: the system stays in state
$\sigma_i$ for $t_a\left(\sigma_i\right)$ time units (until a timeout
event) or until an external event happens, whatever comes first. In
the case of an external event $x_j$, the system changes its state to
$\delta_\mathrm{ext}\left(\sigma_i,e_i,x_j\right)$. In the case of a
timeout, the system changes its state to
$\delta_\mathrm{int}\left(\sigma_i\right)$ and generates an event of
type $\lambda\left(\sigma_i\right)$. In either case, the simulation
time is implicitly advanced to the timestamp of the event that
triggered the transition.  The initial state of the model is not
defined explicitly.

\subsection{Coupled Models}
\begin{figure}[h]\centering
\epsfig{file=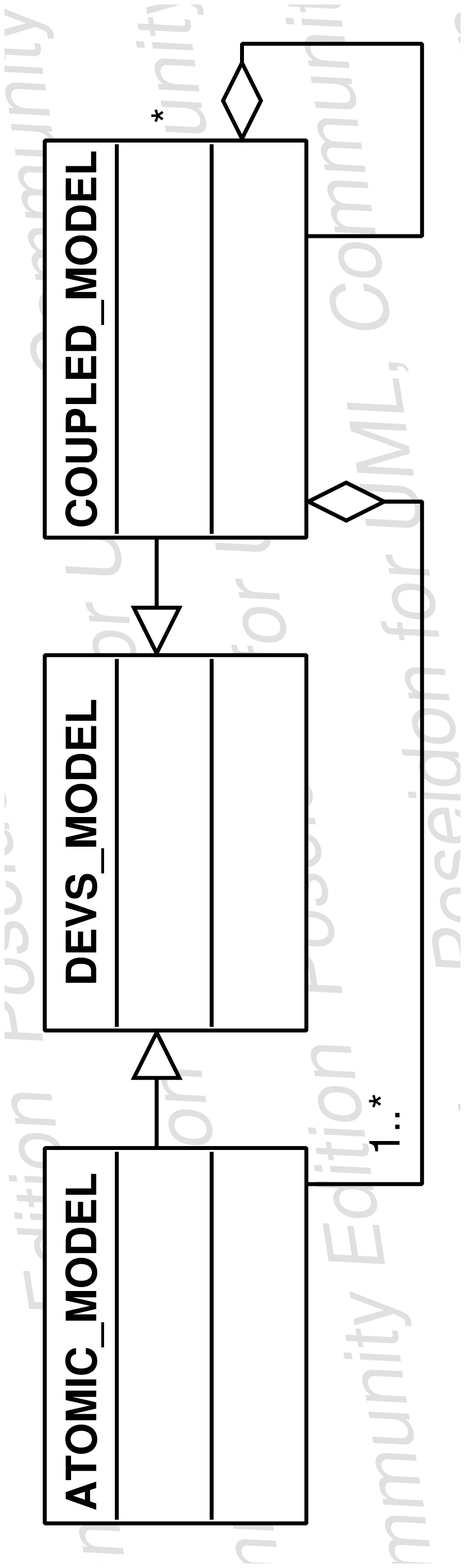,scale=0.3,angle=-90}
\caption{\label{coupled}Atomic and coupled DEVS models}
\end{figure}

Coupled models are used to compose atomic and other coupled models to
produce more DEVS models in a hierarchical way (Figure~\ref{coupled}).

Coupled DEVS model $N$ is a tuple of ten values $N=\{\mathrm{IP},
\mathrm{OP}, X, Y, D, M, E_{IC}, E_{OC}, I_{C}, S\}$. Here,
$\mathrm{IP}$ and $\mathrm{OP}$ are sets of external (not coupled)
input and output ports.  $X$ is a set of input events
$x_i=\left(\pi^{in}_i,v_i\right)$, where $\pi^{in}_i\in IP$ is the
port name, and $v_i$ is the event value, and $Y$ is a set of output
events $y_i=\left(\pi^{out}_i,v_i\right)$, where $\pi^{out}_i\in OP$
is the port name, and $v_i$ is the event value.

$D=\{d_i\}$ is a set of references to the coupled components (atomic
models or other coupled models), and $M=\{M_d|d\in D\}$ is a set of
the coupled components. $E_{IC}\subseteq \{((N, \mathrm{IP}_i), (d,
\mathrm{IP}_{di}))|\mathrm{IP}_i\in \mathrm{IP}, d\in D,
\mathrm{IP}_{di}\in \mathrm{IP}_d\}$ is external input coupling that
connects external input ports of the coupled model to the components'
input ports $\mathrm{IP}_d$. $E_{OC}\subseteq \{(d, \mathrm{OP}_{di}),
((N, \mathrm{OP}_i))|\mathrm{OP}_i\in \mathrm{OP}, d\in D,
\mathrm{OP}_{di}\in \mathrm{OP}_d\}$ is external output coupling that
connects components' output ports $\mathrm{OP}_d$ to the external
output ports of the coupled
model. 

$I_C=\{((a,\mathrm{OP}_{ai}),(b,\mathrm{IP}_{bj}))|a,b\in D,
\mathrm{OP}_{ai}\in \mathrm{OP}_a, \mathrm{IP}_{bj}\in
\mathrm{IP}_b\}$ is internal coupling that interconnects output and
input ports of the components. Finally, $S:\{\epsilon_d|d\in
D\}\rightarrow \epsilon_d$ is a selection function that resolves
potential scheduling conflicts, when more than one event in different
components has the same scheduled time.

Under the principle of closure, a coupled model looks externally  like
an atomic model and can be used anywhere in place of an atomic model.

A coupled model is simulated as an ensemble of its DEVS
components. Output events generated by each individual component are
propagated to the input ports of other coupled components or to the
output ports of the model, according to the functions $I_C$ and
$E_{OC}$. In the former case, they are also converted into appropriate
input events. Input events received by the model are propagated to the
input ports of its components, according to the function $E_{IC}$.

Classic DEVS formalism does not permit feedback loops: $((d, \mathrm{OP}_{di}),
(e, \mathrm{OP}_{ej})) \in I_C \Rightarrow d\ne e$.

\section{UML FORMALISM}

The Unified Modeling Language (UML~\cite{uml}), a de-facto industrial
modeling standard, seems to be a natural choice for a visual
representation of DEVS. Besides being widely supported by both
proprietary and open-source tools (such as Rose~\cite{rose} and
Poseidon~\cite{poseidon}), it also has an associated XML-based
representation, XMI, that makes it possible to process UML diagrams by
application programs.

Of particular interest for us are UML state and component diagrams,
which will be discussed in detail.

\subsection{State Diagrams}

\begin{figure}[bth!]\centering
\epsfig{file=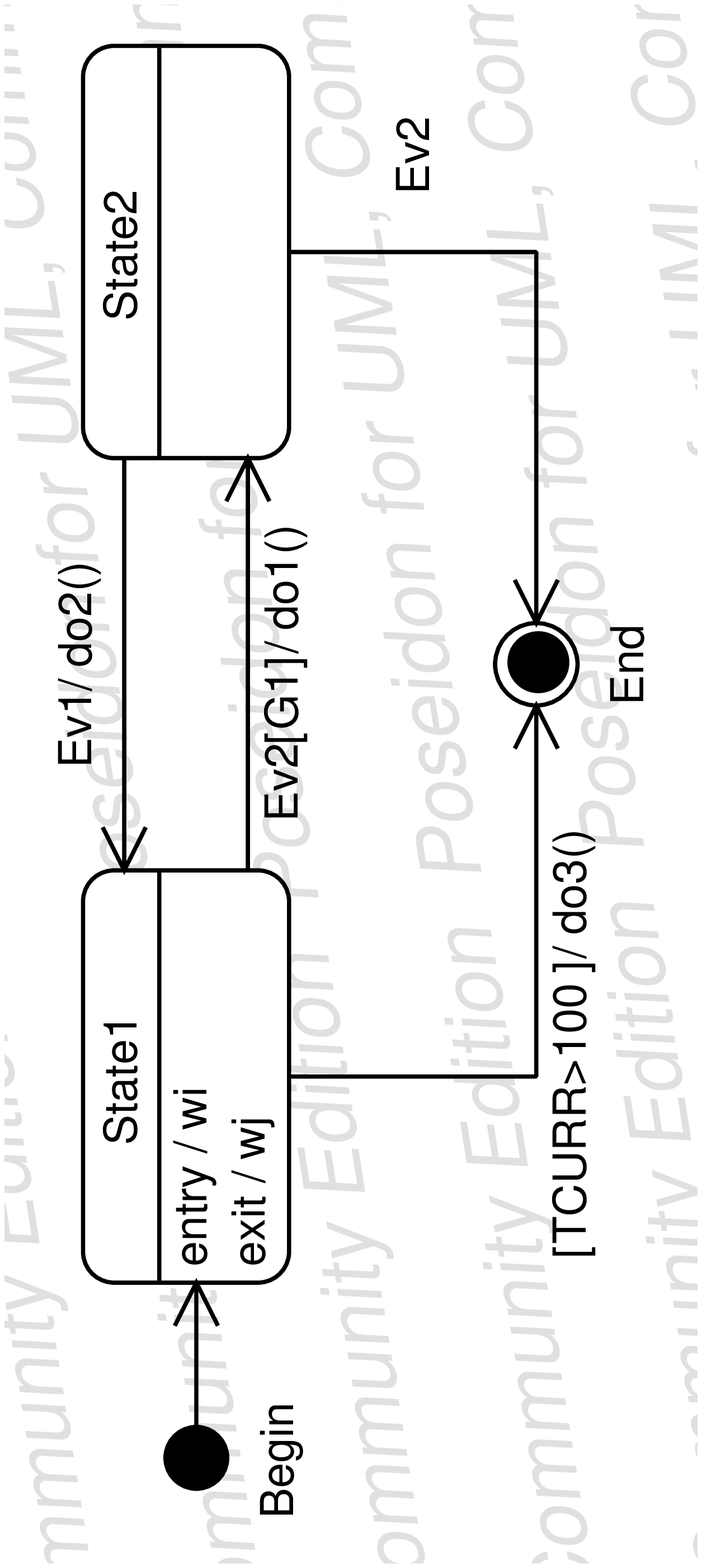,scale=0.3,angle=-90}
\caption{\label{state}A UML state diagram}
\end{figure}

A UML state diagram (also known as a statechart, Figure~\ref{state})
is a visual representation of a finite state automaton with
history. Many of the features of state diagrams, such as ``do''
activities, history states, junction and choice states, concurrent and
composite states, are not essential for DEVS-to-UML mappings and will
not be considered.

A state diagram is a tuple $SD=\{S,S^\bullet,S^\odot,P,T\}$. 

Here $S=\{s_i=(G_i,w_i,q_i)\}$ is a set of finite states. UML finite
states are enumerated using state variable $G$, such that $s_i=s_j
\Leftrightarrow G_i=G_j$. $w_i(x):\mathrm{any}\rightarrow
\mathrm{None}$ is an ``entry'' action. This action is executed just
after changing the current state of the model to
$s_i$. $q_i(x):\mathrm{any}\rightarrow \mathrm{None}$ is an ``exit''
action. This action is executed just before changing the current state
of the model from $s_i$ to some other state.

\begin{figure}[bth!]\centering
\epsfig{file=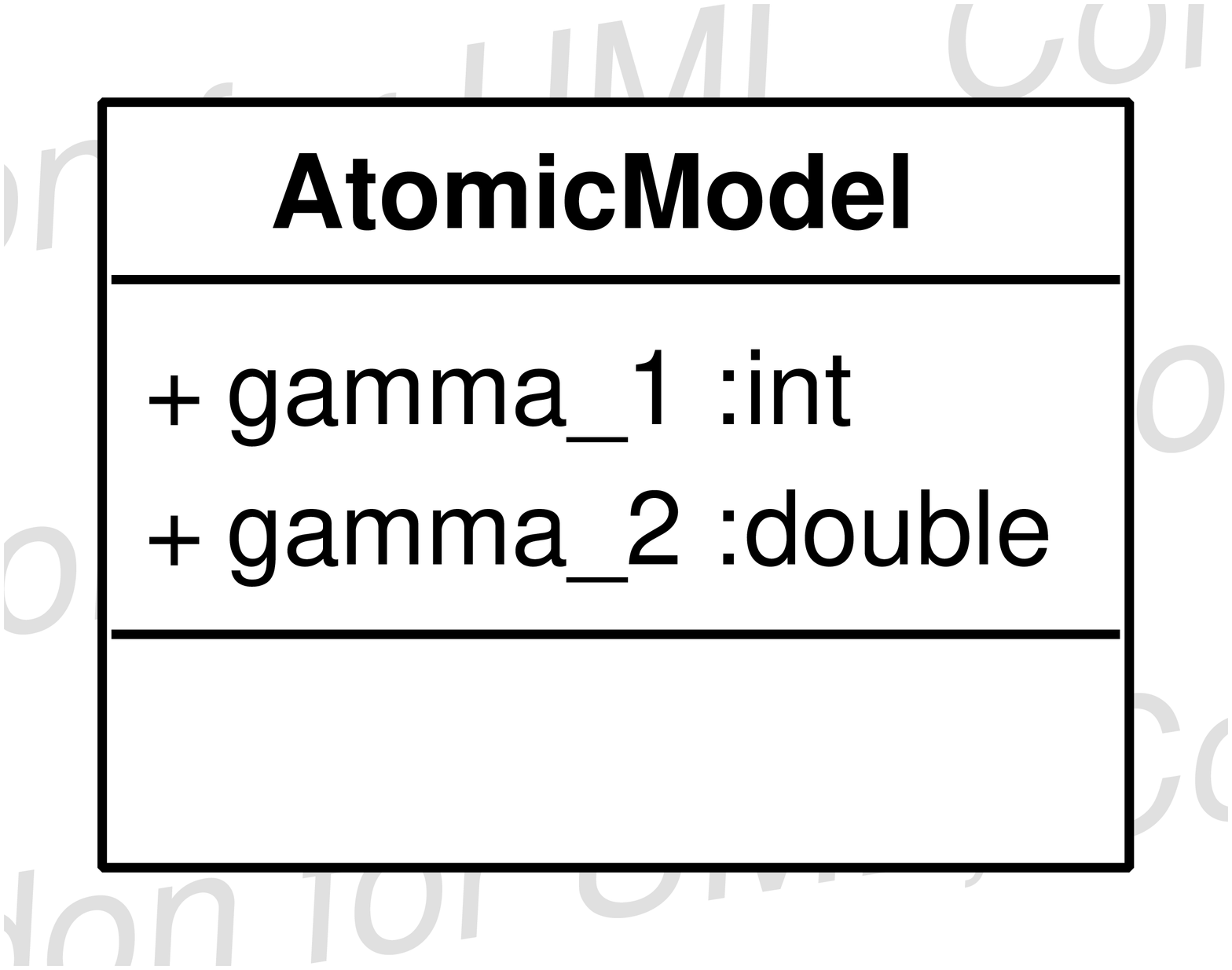,scale=0.15,angle=-90}
\caption{\label{class}A UML class diagram that consists of only one
  class definition. Notice that, formally, variables of types {\tt int}
  and {\tt double} are finite, because both classes have limited range
  and cardinality.}
\end{figure}

A set of possibly continuous {\em pseudostate} variables $H=\{h_k\}$
($|H|\ge 0$) extends the definition of a UML state. The pseudostate
variables may or may not have different values in different
states. They cannot be used to distinguish UML states in a state
diagram. An optional class diagram of the UML system can be used to
record these variables in the model (Figure~\ref{class}).

$S^\bullet=\{s_i\in S\}$ is the set of the initial states of the
diagram. Every diagram can have at most one initial state. Because
DEVS formalism does not specify the initial state of the system, we
will assume that in general $S^\bullet=\emptyset$.

$S^\odot=\{s_i\in S\}$ is the set final (terminal) states of
the diagram. Because the final state of a DEVS system is not defined,
we will assume that in general $S^\odot=\emptyset$.

$P=\{p_j\}$ is a set of discrete events. Each event $p_j$ has the
associated scheduled time $\tau_j$ and a possibly empty set of other
attributes.

Finally, $T=\{s_{bi}, s_{ei}, p_i, g_i, a_i | s_{bi}, s_{ei} \in
S,p_i\in P,g_i\left(x\right):\mathrm{range}(x)\rightarrow
\{\mathrm{True},\mathrm{False}\}, a_i(y):\mathrm{any}\rightarrow
\mathrm{None}\}$ is a set of transitions. In UML notation, a
transition from state $s_{bi}$ to state $s_{ei}$ on event $p_i$ with
guard condition $g_i$ is denoted as $p_i\left[g_i\right]/a_i$. Action
$a_i$ is executed just before the completion of the transition. The
action is not allowed to change the UML state of the system.

The semantics of a UML state diagram prescribes that in the course of
transition $t_i$ from state $s_j$ to state $s_k$ the ``exit'' action
$q_j$ is executed first, followed by the transition action $a_i$,
followed by the ``entry'' action $w_k$.

\subsection{Component Diagrams}
UML component diagrams have evolved substantially from version 1.x of
the language to the current 2.0~\cite{uml,uml04}. The direction of the
evolution has favored the DEVS-to-UML mapping we are about to propose.

In the latest version of the language, deployment, object, and
component diagrams have been merged into a single class of
deployment/object/component (DOC) diagrams. These new DOC diagrams
have a rich language suitable for elaborated models, but for the
purpose of this paper we need to mention only components, interfaces,
and ports.

\begin{figure}[bth!]\centering
\epsfig{file=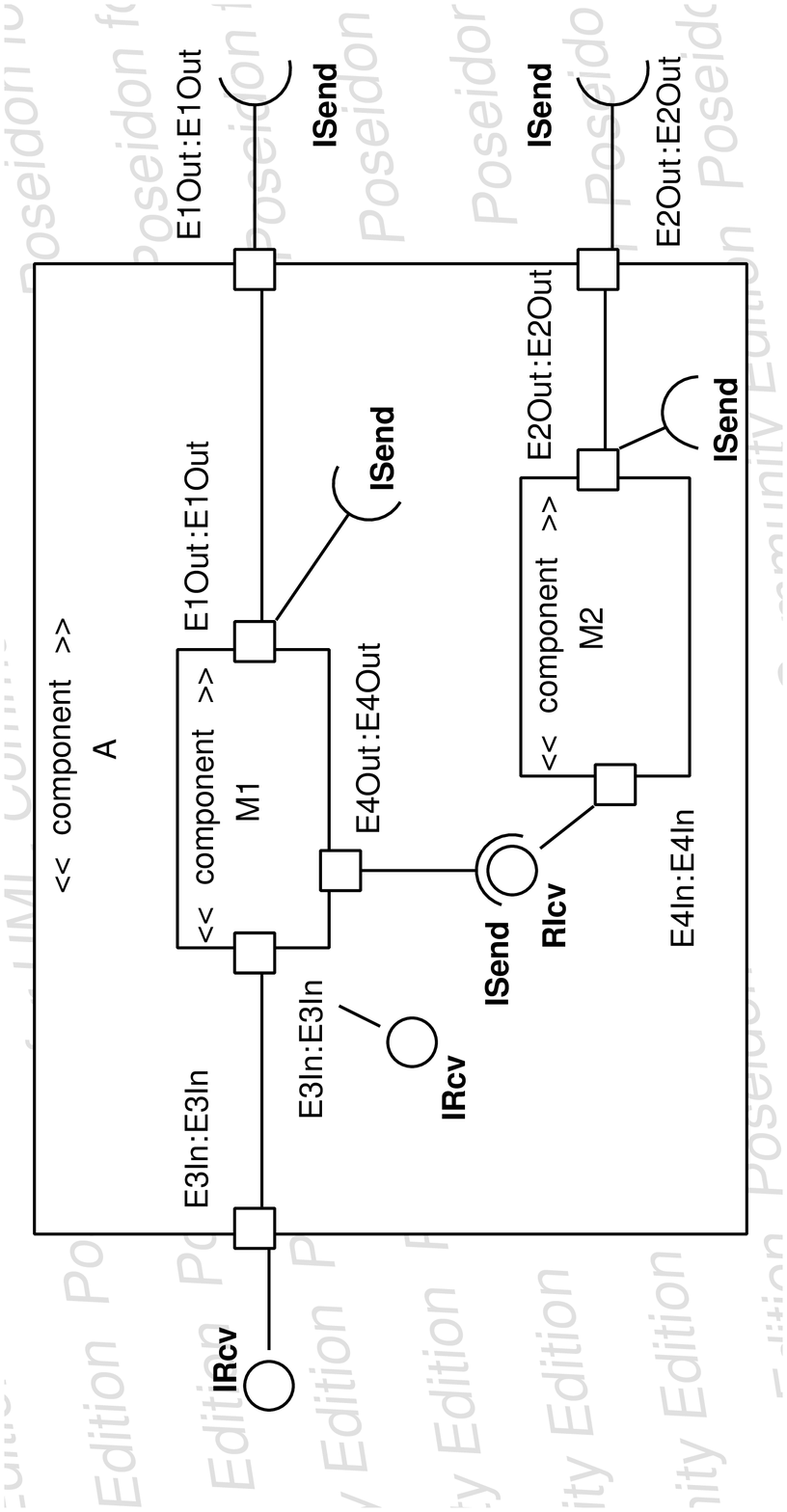,scale=0.3,angle=-90}
\caption{\label{deployment}A UML component diagram}
\end{figure}

A UML component diagram is a tuple $N^u=\{M^u, Q^u, E^u_C, I^u_C\}$.

Here, $M^u=\{M^u_i\}$ is a list of components in the diagram. Each
component $M^u_i$ has list $P^u_i=\{p_{ij}=(t_{ij},if_{ij},d_{ij})\}$
of externally visible ports. A UML port $p$ has a type. The type of
the port specifies the names of the signals (events) that are
acceptable through this port. For the purpose of this paper, we assume
that the type $t$ of the port is the same as its name. A port can be
unidirectional (input or output) or bidirectional. The direction of
the port $\mathrm{dir}(p_{ij})$ is defined by the type(s) of its
interfaces.  Since DEVS formalism does not support bidirectional
ports, we will not consider them and consider only one interface
$K=(n,d)$ per port, where $n$ is the name of the interface and
$d=\{\mathrm{required}|\mathrm{provided}\}$ is its type. Required
interfaces (``antennas'') define output ports, and provided interfaces
(``lollipops'') define input ports.

A UML component represents either another UML component diagram, or
a UML state diagram.

Set $Q^u \subseteq \{(M^u_i, M^u_j)|M^u_i\subset M^u\wedge
M^u_j\subset M^u\}$ defines the containment relation: component $M$ is
a subcomponent of component $N$, or a nested component, if $(N,M)\in
Q^u$. Let $Z(N)=\{M_i|M_i\in M^u\wedge ((N,M_i)\in Q^u \vee (\exists
M_j: M_j\in Z(N)\wedge (M_j,M_i)\in Q^u)\}$ be the set of {\em
children} of $N$. The containment relation must satisfy additional
conditions: (a) a component cannot be a subcomponent of itself:
$\forall M\in M^u: (M,M)\not\in Q^u$ and (b) if $M$ is a subcomponent
of component $N$, then $N$ cannot be a subcomponent of $M$ or of any
child of $M$: $(N,M)\in Q^u\leftrightarrow (M,N)\not\in Q^u \wedge
\forall M'\in Z(M):(M',N)\not\in Q^u$.

Components $T_{N^u}=\{T_i|T_i\in M^u\wedge\forall M\in
M^u:(T_i,M)\not\in Q^u\}$ are called top-level components. The  set:

\begin{equation}
\begin{split}
E^u_C&\subseteq \{((M_i, p_{ik}),M_j, p_{jl})) |\\
&M_i\in M^u\wedge M_j\in M^u\wedge \\
&(M_i,M_j)\in Q^u\wedge \\
&p_{ik}\in P^u_i\wedge p_{jl}\in P^u_j\wedge\\
&\mathrm{dir}(p_{ik})=\mathrm{dir}(p_{jl})\}
\end{split}
\end{equation}
\noindent%
is the external coupling that connects external ports of the component to
the subcomponents' ports using UML {\em delegation
connectors}. External ports must be connected to the internal ports of
the same direction.
The set:
\begin{equation}
\begin{split}
 I^u_C&\subseteq \{((M_i, p_{ik}), (M_j, p_{jl})) |\\
& M_i\in M^u \wedge M_j\in M^u\wedge\\
&\exists M_p\in M^u:\{(M_i,M_p),(M_j,M_p)\}\subseteq Q^u\wedge\\
&p_{ik}\in P^u_i\wedge p_{jl}\in P^u_j\wedge\\
&\mathrm{dir}(p_{ik})\ne\mathrm{dir}(p_{jl})\}
\end{split}
\end{equation}
\noindent%
is the internal coupling that interconnects ports of the subcomponents of
the same component using UML {\em assembly connectors}. Internal ports
must be connected to the internal ports of the opposite
direction. Graphically, this is accomplished by matching ``antennas''
and ``lollipops''.

A simple UML component diagram (strictly speaking, a DOC diagram) is
shown in Figure~\ref{deployment}. It depicts two components M1 and
M2. These components are in turn subcomponents of component
A. Component A has two output ports E1Out and E2Out with interface
ISend and one input port E3In with interface IRcv. The output ports of
subcomponents M1 and M2 are connected to the corresponding output
ports of the component A, the input port of A is connected to the
input port of subcomponent M1. Finally, the output port E4Out of
component M1 with interface ISend is connected to the input port E4In
of component M2 with interface IRcv.

Composite structure diagrams, which only recently have become a part
of the UML 2.0 proposal~\cite{hogg04}, offer even better mapping
between DEVS models and UML models. However, the discussion of these
diagrams is beyond the scope of this paper.

\section{ATOMIC DEVS AND UML STATE DIAGRAMS}
To map an atomic DEVS model into an UML state diagram, we need to map
states, ports, transitions, and outputs.

\subsection{States}
In general, DEVS states are neither discrete nor finite, while UML
states are always discrete and finite. To map a DEVS system onto an
UML system, the DEVS state must be discretized. This can be done by
rearranging and grouping DEVS state variables $\gamma=\{\gamma_j\}$
according to their kind. Discrete and finite variables can be
collected in one subgroup, and countable and continuous
variables---into another subgroup:

\begin{equation}
\begin{split}
\gamma=(\underbrace{\gamma_1,\ldots,\gamma_m},\label{eq:partition}
\underbrace{\gamma_{m+1},\ldots,\gamma_n}),& 1\le m\le n\\
|\mathrm{range}(\gamma_j)|<\infty,& 1\le j\le m\\
|\mathrm{range}(\gamma_j)|=\infty,& m< j\le n\\
\end{split}
\end{equation}

Let's call the first group {\em finite} DEVS state variables, and
the other group {\em free} DEVS state variables.

For any feasible combination of values of finite DEVS state variables
$\{\gamma_i|1\le i\le m\}$, a UML finite state can be constructed by
simply enumerating this combination: $G_j=\bigotimes_{i=1}^m\gamma_i$,
where the details of the function $\bigotimes$ are not important as
long as it always maps distinct combinations of finite DEVS state
variables onto distinct UML states.

All remaining DEVS free state variables $\{\gamma_i|m< i\le n\}$ are
mapped to UML pseudostate variables $H=\{h_i|1\le i< n-m\}$
one-to-one. They become attributes of finite states and will be
manipulated during DEVS state transitions.

In the case of $m=0$ a DEVS model has no finite state variables, and
partitioning (\ref{eq:partition}) is not possible. The UML state
diagram will have only one final state. This case is presented
in~\cite{feng04}. 

On the other hand, when $m=n$, the DEVS model has no free
variables. Every feasible combination of the DEVS state variables maps
onto a UML finite state, and the UML model has no pseudostate
variables.

Not being a dynamic simulation language, UML does not enforce an
explicit notion of time in all types of diagrams. In particular, UML
state diagrams do not have explicit time. The simulation time has to
be maintained implicitly, with model-wide variable $t_\mathrm{curr}$ denoting the
current simulation time.

For the purpose of computing state transitions, many DEVS models
depend on the time $e_i$ spent by the model in state $\sigma_i$. To
accommodate this need, we introduce another model-wide variable
$t_e$---the time of the most recent state transition. An ``entry''
action $w_i$ can be added to every UML finite state that records the
value of $t_e$:

\vskip\baselineskip
\noindent%
\strut\hspace{0.2in}def $w_i$ (): \\
\strut\hspace{0.4in}$t_e=t_\mathrm{curr}$
\vskip\baselineskip

At any given time, the value of $e_i$ can be now computed as:

\begin{equation}
\label{ei}
e_i=
\begin{cases}
  t_\mathrm{curr}-t_e&\text{if $i$ is the current state,}\\
  +\infty&\text{otherwise.}
\end{cases}
\end{equation}

\subsection{Ports}
DEVS input and output ports are mapped to UML events one-to-one:

$$
\forall \pi_i\in \left(\mathrm{IP}\cup \mathrm{OP}\right)
\exists !p_j\in P.
$$
Input and output events are not distinguished in UML state diagrams.

DEVS formalism does not specify whether the same port can be used as
an input port and an output port (whether $\mathrm{IP}\cap
\mathrm{OP}=\emptyset$). We will assume that the DEVS ports are indeed
unidirectional. This means that an event of a certain type can be only
consumed or produced by a state diagram, but not both.

\subsection{External Transitions}
The heterogeneous structure of DEVS states and UML states produces
heterogeneous structures of DEVS and UML state transitions. Among
other things, a single DEVS external transition function
$\delta_\mathrm{ext}\left(\sigma_i,e_i,x_j\right)=(\pi^{in}_j,\nu_j)$ has
to generate both proper UML state transitions and their associated
guard conditions and actions.

The total state $S_{T_i}$ of a UML model is defined by the current UML
finite state and the values of the pseudostate variables:
$$
S_{T_i}\equiv\left(s_i, H\right), |H|=n-m+1.
$$ 

The total state $\Sigma_i$ of a DEVS model is defined by the values of
all DEVS state variables:
$$ \Sigma_i\equiv\left(\gamma_{i1}, \ldots,
\gamma_{in}\right)=\gamma_i.
$$

To reflect a transition from a DEVS state $\sigma_i$ to another state
$\sigma_j$, the corresponding UML model has to change from a UML
state $s_i=f(\gamma_i)$ to another UML state $s_j=f(\sigma_j)$ and
also update the values of the pseudostate variables:
$H_j=f(\gamma_j)$. Here, $f(\chi):\mathrm{DEVS\_objects}\rightarrow
\mathrm{UML\_objects}$ is the polymorphic mapping function from the
universe of DEVS objects to the universe of UML objects which has been
partially defined above.

Action $a^e_l$ associated with a possible UML transition
$t_{l=(ijk)}=\{s_i,s_j,p_k,g^e_l,a^e_l\}$ from state $s_i$ to state $s_j$
on event $p_k=f(\pi^{in}_k)$ would be responsible for updating the
values of pseudostate variables $H$. It may depend on the original
values of the pseudostate variables, on the value of the event, and on
$e_i$:

\vskip\baselineskip
\noindent%
\strut\hspace{0.2in}def $a^e_l$ ($H_i$, $\nu_k$, e): \\
\strut\hspace{0.4in}$H_j$=$z^e_l$($H_i$, $\nu_k$, e)
\vskip\baselineskip

Here, $z^e_l(H,\nu,e)$ is the explicit state update function:
\begin{equation}
\begin{split}
z^e_l (H,\nu,e)&=H'\leftrightarrow\\
&e\ge 0 \wedge\\
& \delta_\mathrm{ext}\left(\left(s_i,H\right),e,(\pi^{in}_k,\nu)\right)
=\left(s_j,H'\right).
\end{split}
\end{equation}

This function may be undefined for some or all values of $H$, $\nu$,
 and $e=e_i$. A {\em condition} for a possible UML transition is set
 $C^e_l$ constructed in the following way:

\begin{equation}
C^e_l=\{\left(H,\nu,e\right)|(H,\nu,e)\in\mathrm{dom}(z^e_l)\}.
\end{equation}
 
The UML transition exists if and only if the corresponding
condition is not an empty set:

$$ t_l\in T\leftrightarrow C^e_l\ne\emptyset.
$$

Note that multiple transitions from $s_i$ to $s_j$ may exist for
different events $x_k$. In general, $s_i$ can be the same state as
$s_j$ (loopback transitions are possible).

The guard condition $g_{k}$ is a functional representation of the
condition set $C^e_l$:

\begin{equation}
g^e_l(H,\nu,e)=
\begin{cases}
\mathrm{true}&(H,\nu,e)\in C^e_l,\\
\mathrm{false}&\text{otherwise},
\end{cases}
\end{equation}
\noindent%
where $e$ for state $s_i$ is defined by Eq.~\ref{ei}. The transition is
``fired'' if event $p_k$ occurs and $g^e_l(H,\nu_k,e)$ is true.

\subsection{Internal Transitions and Output Events}
Internal transitions and generation of output events are regulated by
the internal transition function
$\delta_\mathrm{int}\left(\sigma_i\right)$, the time advance function
$t_a\left(\sigma_i\right)$, and the output function
$\lambda\left(\sigma_i\right)$. The three functions co\"operate in the
sense that the first function computes the target state of the
transition, the second schedules the transition, and the third
generates an output event associated with the transition. DEVS models
allow to output events only during internal transitions.

As the external UML transitions, internal UML transitions need to
advance the UML model to another state and to update the pseudostate
variables. 

Action $a^i_l$ associated with a possible UML transition
$t_{l=(ijk)}=\{s_i,s_j,\epsilon,g^i_l,a^i_l\}$ from state $s_i$ to state
$s_j$ on {\em null} event $\epsilon$ would be responsible for updating
the values of pseudostate variables $H$. It may depend on the original
values of the pseudostate variables. 

Unlike an action associated with an external transition, $a^i_l$ can
generate output events.  UML2.0 allows UML components to ``send''
events to any UML component $M$, including the component that sends
the event (``self''), either immediately, or later. Keyword ``after''
that is used to schedule an event in the future. Caret (\^{})
represents the send operation. For example, \underline{\^{}M.e
after $t$} means ``send event $e$ to
component $M$ after time $t$.''

The function $\lambda\left(\sigma_i\right)$ defines whether an output
event is generated during the transition or not, and if yes, what is
the type (port) of the event $\pi^{out}_i$ and its value $\nu_i$. The
value of the event can be stored in the UML model as its attribute. By
construction, the name of the output event and the name of the output port of
an atomic DEVS model coincide:

\vskip\baselineskip
\noindent%
\strut\hspace{0.2in}def $a^i_l$ ($H_i$): \\
\strut\hspace{0.4in}($\pi^{out}_i,\nu_i$)=$\lambda\left(\sigma_i\right)$\\
\strut\hspace{0.4in}\^{}$\pi^{out}_i.\left(\pi^{out}_i(\nu_i)\right)$ \\
\strut\hspace{0.4in}$H_j$=$z^i_l$($H_i$)
\vskip\baselineskip

Here, $z^i_l(H)$ is the explicit state update function for internal
transitions:
\begin{equation}
z^i_l (H)=H'\leftrightarrow \delta_\mathrm{int}\left(\left(s_i,H\right)\right)
=\left(s_j,H'\right).
\end{equation}

This function may be undefined for some or all values of $H$. A {\em
 condition} for a possible internal UML transition is set $C^i_l$
 constructed in the following way:

\begin{equation}
C^i_l=\{H|H\in\mathrm{dom}(z^i_l)\}.
\end{equation}
 
The internal UML transition exists if and only if the corresponding
condition is not an empty set:

$$ t_l\in T\leftrightarrow C^i_l\ne\emptyset.
$$

In general, $s_i$ can be the same state as $s_j$ (internal loopback
transitions are possible).

The guard condition $g_{k}$ is a functional representation of the
condition set $C^i_l$:

\begin{equation}
g^i_l(H)=
\begin{cases}
\mathrm{true}&H\in C^i_l,\\
\mathrm{false}&\text{otherwise},
\end{cases}
\end{equation}

\noindent%
The time spent in DEVS state $\sigma_i$ before the transition is
scheduled, is given by the function $t_a\left(\sigma_i\right)$. 

It is tempting to use the send/after apparatus to schedule the
transition in the future. However, a transition is not an event and
cannot be scheduled using ``after'' and ``send''. Instead, we declare
new {\em timeout} event $e_i$ and schedule it at time
$t_\mathrm{curr}+t_a((s_i,H_i))$ be redefining the entry action $w_i$ of state
$s_i$.

Event $e_i$ becomes the trigger for the internal transition from state
$s_i$. 

A situation may occur when a timeout event has been scheduled, and an
external transition is triggered by another (input) event. In this
case, the pending timeout must be canceled. UML2.0 does not allow to
recall scheduled events. Instead, a guard condition can be changed to
make sure that obsolete timeouts do not trigger internal transitions.

Let every UML state $s_i$ have local variable $y_i$ initialized to the
reference to the most recently scheduled timeout event $e_i$ in the
entry action:

\vskip\baselineskip
\noindent%
\strut\hspace{0.2in}def $w_i$ (): \\
\strut\hspace{0.4in}$t_e=t_\mathrm{curr}$\\
\strut\hspace{0.4in}$y_i=\mathrm{ref}(e_i)$\\
\strut\hspace{0.4in}\^{}self.$e_i$ after $t_a((s_i,H_i))$
\vskip\baselineskip

If the value of $y_i$ and the reference to the scheduled timeout
$e_i$ differ, the timeout event must be ignored.

To summarize, an internal transition from state $s_i$ is scheduled
when $e_i$ occurs and if $\left(y_i==\mathrm{ref}(e_i)\right)\wedge
g^i_l(H)$ is true.

\section{COUPLED DEVS AND UML COMPONENT DIAGRAMS}

DEVS coupled models and UML component diagrams have a very similar
structure, which makes mapping DEVS models onto UML component diagrams
rather straightforward.

\subsection{Components}

A DEVS coupled model $N$ has only one top-level component and only one
level of nesting. This corresponds to a UML component diagram $N^u$
with $T_N=\{T_0\}$ and such that $\forall M_i\in M^u\setminus
T_N,\forall M_j\in M^u:(M_i,M_j)\not\in Q^u$.

The top-level component $T_0$, therefore, represents the DEVS coupled
model itself, and for each DEVS component $M_d\in M$ there exists an
UML component $M^u_i\in M^u,M^u_i=f(M_d)$. 

\subsection{Ports}
\setlength{\textheight}{7.5in}
Both input ports $\mathrm{IP}$ and output ports $\mathrm{OP}$ of $N$
are mapped to the corresponding UML ports of the top-level component
$T_0$ with externally visible ports $P^u_0$. Let port
$p_i\in\mathrm{IP}$ carry input events of type $x_i\in X$. Then for
this port $\exists p^u_i=f(p_i)=(t,i\!fc,d)\in P^u_0:d = \mathrm{input}\wedge
t=x_i$. Let port $p_j\in\mathrm{OP}$ carry output events of type
$y_j\in Y$. Then for this port $\exists p^u_j=f(p_j)=(t,i\!fc,d)\in P^u_0:d =
\mathrm{output}\wedge t=y_j$. Notice that input and output events are
mapped to the port names implicitly.
 
\subsection{Connectors}
External DEVS coupling corresponds to external (delegation) UML
connectors. External input coupling of input port $p_i\in \mathrm{IP}$
to an input port $p^u_j=f(p_i)$ of component $M_d\in M$ is mapped to a
delegation connector from the corresponding port of $T^u_0$ to the
corresponding port of $M^u_i$: $\forall c_i=((N, \mathrm{IP}_i), (d,
\mathrm{IP}_{di}))\in E_{IC}\exists c^u_j=((T_0, p_{0j}),(M^u_k,
p_{kl})): M^u_k=f(M_d)\wedge p_{0j}=f(\mathrm{IP}_i)\wedge
p_{kl}=f(\mathrm{IP}_{di})$. Respectively, external output coupling of
an output port $p^u_j=f(p_i)$ of component $M_d\in M$ to output port
$p_i\in \mathrm{OP}$ is mapped to a delegation connector from the
corresponding port of $M^u_i$ to the corresponding port of $T^u_0$:
$\forall c_i=((d, \mathrm{OP}_{di}), (N, \mathrm{OP}_i))\in
E_{OC}\exists c^u_j=((M^u_k, p_{kl}),(T_0, p_{0j})):
M^u_k=f(M_d)\wedge p_{0j}=f(\mathrm{OP}_i)\wedge
p_{kl}=f(\mathrm{OP}_{di})$.

Internal DEVS coupling corresponds to internal (assembly) UML
connectors. Internal coupling of port $p^u_j=f(p_i)$ of component
$M_d\in M$ to port $p^u_l=f(p_k)$ of component $M_e\in M$ is mapped to
an assembly connector from the corresponding port of $M^u_i=f(M_d)$ to
the corresponding port of $M^u_k=f(M_e)$.

\subsection{Selection function}
There is no feasible way of mapping the DEVS selection function to a
UML model. Fortunately, parallel DEVS models do not use this function
at all.

\section{CONCLUSION AND FUTURE WORK}
In the paper, we proposed a mapping of discrete event specification
(DEVS) models onto Unified Modeling language (UML) state and component
diagrams. This diagrams are designed and optimized for computerized
processing and are highly expressive, competitive, and widely used
both in academia and industry. Successful automated DEVS-to-UML
mapping techniques would enable seamless integration of the legacy of
DEVS models into existing and emerging UML models.

As the future step, we plan to consider UML2.0 composite structure
diagrams as more suitable representations of DEVS coupled
models. An automated translator from the DEVS domain into the
UML domain would substantially simplify the transformation. Limiting
the output to the Executable UML is also considered.

\section*{ACKNOWLEDGMENTS}
The author is grateful to Prof.~T.~G.~Kim of KAIST and Dr.~M.-H.~Hwang
of VMS Solutions for their inspiring discussions and suggestions
during the Summer Computer Simulation Conference'2004. Positive,
helpful, and friendly input from Prof.~P.~Ezust and
Prof.~D.~\c{S}tefan\u{a}scu of Suffolk University substantially improved the
paper. Continuous interaction with A.~Al-Shibli of Suffolk University
enhanced my understanding of UML diagrams and helped me to avoid many common
mistakes. 

\section{BIOGRAPHY}

Dr. D.~Zinoviev received his Ph.D. in Computer Science from SUNY at
Stony Brook in 1997. He was working as a post-doc on the
DARPA/NASA/NSA-sponsored Petaflops project of a hybrid technology,
multi-threaded hypercomputer. In 2000, he joined the Computer Science
Department of Suffolk University in the rank of Assistant
Professor. His current research interests include simulation and
modeling (network simulation, architectural simulation), operating
systems, and software engineering.

\bibliographystyle{plain}
\bibliography{paper}

\begin{thebibliography}{10}

\bibitem{uml}
G.~Booch, J.~Rumbaugh, and I.~Jacobson.
\newblock {\em The Unified Modeling Language. User Guide}.
\newblock Addison Wesley, 1999.

\bibitem{feng04}
H.~Feng.
\newblock {DCharts}, a formalism for modeling and simulation based design of
  reactive software systems.
\newblock Master's thesis, School of Computer Science, McGill University,
  Montr\'eal, Canada, February 2004.

\bibitem{poseidon}
GentleWare.
\newblock Poseidon for {UML} community edition v. 2.5.1.
\newblock Available at http://gentleware.com.

\bibitem{hogg04}
J.~Hogg.
\newblock {UML} 2.0 automates code generation from architecture.
\newblock {\em {COTS} Journal}, May 2004.
\newblock Available online at http://www.cotsjournalonline.com.

\bibitem{hong04}
S.-Y. Hong and T.~G. Kim.
\newblock Embedding {UML} subset into object-oriented {DEVS} modeling process.
\newblock In A.~Bruzzone and E.~Williams, editors, {\em Proc. SCSC 2004}, pages
  161--166, San Jose, CA, July 2004.

\bibitem{rose}
IBM.
\newblock Rational rose.
\newblock Available at \\http://ibm.com/software/rational/.

\bibitem{mellor2002}
S.~J. Mellor and M.~J. Balcer.
\newblock {\em Executable {UML}: A Foundation for Model Driven Architecture}.
\newblock Addison-Wesley, 2002.

\bibitem{uml04}
J.~Rumbaugh, I.~Jacobson, and G.~Booch.
\newblock {\em The Uni\-fied Modeling Language Reference Manual}.
\newblock Addison Wesley, second edition, 2004.

\bibitem{sarjoughian04}
H.~Sarjoughian and R.K. Singh.
\newblock Building simulation modeling environments using systems theory and
  software architecture principles.
\newblock In {\em Proc. Advanced Simulation Technology Symposium}, Washington
  DC, April 2004.

\bibitem{schulz00}
S.~Schulz, T.C. Ewing, and J.W. Rozenblit.
\newblock Discrete event system specification ({DEVS}) and {StateMate}
  {StateCharts} equivalence for embedded systems modeling.
\newblock In {\em Proc. 7th IEEE International Conference and Workshop on the
  Engineering of Computer Based Systems}, pages 308--316, April 2000.

\bibitem{zeigler84}
B.P. Zeigler.
\newblock {\em Multifaceted Modelling and Discrete Event Simulation}, chapter
  4, 7.
\newblock Academic Press, 1984.

\bibitem{zeigler90}
B.P. Zeigler.
\newblock {\em Object-oriented Simulation with Hierarchical, Modular Models},
  chapter~3.
\newblock Academic Press, 1990.

\bibitem{zeigler00}
B.P. Zeigler, H.~Praehover, and T.G. Kim.
\newblock {\em Theory of Modeling and Simulation}, chapter~4.
\newblock Academic Press, 2000.

\end{thebibliography}

\end{document}